\documentclass[manuscript,screen]{acmart}
\AtBeginDocument{%
  }

\begin{document}

\title{Sustainable and Adaptive Growth in Computing Education} 

\author{Enes Ayalp}
\email{Anonymous}
\email{enes.ayalp@gmail.com}
\affiliation{%
  \institution{Georgia Institute of Technology}
  \city{Atlanta}
  \state{}
  \country{United States}
}


\renewcommand{\shortauthors}{Enes Ayalp}

\begin{abstract}
Computing Education faces significant challenges in equipping graduates with the resilience necessary to remain relevant amid rapid technological change. While existing curricula cultivate computing competencies, they often fail to integrate strategies for sustaining and adapting these skills, leading to reduced career resilience and recurrent industry layoffs. 
The lack of educational emphasis on sustainability and adaptability amid industry changes perpetuates a vicious cycle: As industries shift, skill fragmentation and decay lead to displacement, which in turn causes further skill degradation. The ongoing deficiency in adaptability and sustainability among learners is reflected in the frequent and intense shifts across the industry.
This issue is particularly evident in domains marked by high technological volatility such as computer graphics and game development, where computing concepts— including computational thinking and performance optimization—are uniquely and continuously challenged. 
To foster sustainable and adaptive growth, this paper introduces, a new framework which addresses the question: How can computing education and professional development be connected to in these volatile sectors?
It integrates two iterative, interconnected cycles—an educational and a professional—by linking education with profession to establish a lifelong, renewable practice. This approach allows computing professionals to excel and maintain relevance amid constant changes across their industry.

\end{abstract}





\maketitle

\section{Introduction}
Individuals who combine competencies in computer science with creative practice continue to be highly sought after, especially in creative technology domains—such as gaming and film \cite{tiga2024, combine2025, shijia2023}—which are among the most competitive and volatile professional domains. Despite continued industry growth, many of them shift roles or exit the field altogether due to career instability and uncertainty. Some anticipate long-term unemployment and seek greater stability in other industries, while others, facing layoffs, feel compelled to leave competitive sectors \cite{sharma2019, crevoshay2019, igda2024}. This pattern raises a critical question: why do industries that depend on computer science struggle to retain even most capable professionals?

In fields such as computer graphics and game development, professionals consistently demonstrate exceptional technical and creative abilities involving computational thinking, evident through platforms like ArtStation \cite{artstation2022} and game jams \cite{globalgamejam2025}. With numerous avenues for ongoing collaboration, learning and improvement, their challenges—and those of the industry—cannot be addressed through technical proficiency alone.
Because education’s role extends beyond skill acquisition to preparing learners as industry-ready professionals, it is key to examine which domains face the greatest challenges in transforming students into professionals, and which areas of education are most responsible for enabling that transformation.

Creative-technology, a subdomain of both computer science and art refers to fields that combine computing with creative expression such as game development, computer graphics, interactive media and digital art. The fields in this domain place a greater emphasis on computing education compared to other disciplines (e.g. art), because computing technologies evolve more rapidly and their changes have a much bigger impact on industry practices. While artistic creations often remain timeless, computing tools and frameworks can become obsolete within years, rendering previously valuable expertise less relevant. 

Creative technologists—a group of passionate professionals depending heavily on computing education—are expected to possess both deep specialization and broad applicability yet pursuing both simultaneously overextend their resources, making progress in either direction less efficient \cite{kelly2011}. Given the highly competitive nature of their industry, even a brief period of skill irrelevance can prompt a career shift or result in unemployment. These professionals are expected to excel creatively and technically while understanding how their work fits into the broader pipeline, collaborating effectively, and adapting to new tools and workflows \cite{sosa2022}. The tension between depth and breadth leaves practitioners in a precarious position: deep specialists risk obsolescence when tools or workflows change \cite{mcguinness2023}, whereas generalists struggle to distinguish themselves in highly competitive markets \cite{moeran2010}.

Acquiring computer science skills is increasingly accessible through diverse pathways, including fully independent learning; however, sustaining and adapting these abilities into stable careers remains a persistent challenge especially in creative technology \cite{comunian01122011}. Investments in learning often exceed the returns \cite{fcomm2024}, and "there is substantial skill loss with non-practice or non-use" \cite{arthur1998}. Empirical studies suggests that proactive engagement can enhance career outcomes \cite{chang2023}, and this is widely promoted within creative industry. Yet, when these strategies were widely adopted, competition intensified and returns remained limited, leaving even highly skilled individuals with persistent instability—a reflection of the unequal rewards for professional initiative.

Sustaining professionally in the industry requires visibility (a typical form of proactive engagement), yet research over the past decade indicates that the focus on personal branding often introduced more complications than it supported growth. \cite{intmar2011, mei2024}.
To address these gaps, a conceptual, structured framework is designed that guides creative technologists in their learning and professional activities, helping them navigating through the known industry challenges. The goal is to support practitioners in building stable, resilient careers while continuing to pursue their continuous education in computing. By effectively supporting creative technologists, who exemplify the challenges inherent in computing education, this approach can extend its benefits to other computing roles, since the current gaps affect the entire computer science domain.

\section{Framework Requirements}
To address the previous mentioned challenges, the proposed framework must satisfy a set of essential requirements that ensure both professional growth and career sustainability. The following requirements reflect the need to cultivate technical and creative competence, integrate into professional contexts, maintain long-term relevance, and sustain intrinsic motivation:

\begin{enumerate}
\item \textbf{Skill Evaluation} ensures technical and creative competence to meet professional standards.
	
\item \textbf{Credibility} demonstrates recognizable competence, reliability, and value.

\item \textbf{Sustainability} ensures long-term career and financial viability amid industry change.

\item \textbf{Adaptability} is key to maintain relevance in any industry.  

\item \textbf{Passion} supports engagement, originality, and long-term commitment to creative practice. 

\item \textbf{Versatility} enables the contribution across multiple domains using transferable specialized skills. 
\end{enumerate}

These requirements must be satisfied to cultivate technical and creative excellence, ensure professional recognition, and support sustainable, adaptable, and passion-driven careers in the competitive field of creative technology.

\section{Mapping Known Methods}

Existing methods are examined to identify how effectively they satisfy the previously defined requirements. The framework is then designed with the aim of achieving outcomes that remain relevant over time.
To comparatively assess these methods, a structured qualitative evaluation was conducted. Each method was analyzed against the six requirements defined in the previous section. Rather than relying on survey data, this evaluation is heuristic in nature — based on theoretical reasoning, professional observation, and comparative strengths identified in literature and practice. Each method was scored on a scale from 0 (not promoting) to 10 (strongly promoting) for each requirement. These scores serve as conceptual indicators of relative emphasis, rather than quantitative measures of performance.

Following this evaluation design, each method was examined individually to justify its assigned scores. The reasoning for these scores is based on the observed alignment of each method with the six framework requirements—namely, \textit{Skill Evaluation}, \textit{Credibility}, \textit{Sustainability}, \textit{Adaptability}, \textit{Passion}, and \textit{Versatility}. The qualitative reasoning presented below outlines how each method promotes these requirements and explains the rationale behind the values reported in Table~\ref{tab:methods_scores}.

\subsection{Qualitative Evaluation of Existing Methods Against Defined Requirements}

\subsubsection{Game Jams}
\begin{itemize}
	\item Skill Evaluation (8): High due to multiple roles and tight deadlines \cite{8686138}; slightly reduced because teamwork can buffer individual skill pressure \cite{saud2024}.
	
	\item Credibility (7): Project outcomes are publicly visible and provide networking opportunities but they depend on the ability to effectively collaborate with peers \cite{10.1145/3196697.3196700}.
	
	\item Sustainability (3): Occasional transition to commercial projects exists, but most jams do not generate predictable income; mainly a learning and networking platform \cite{10.1145/2677758.2677778}.
	
	\item Adaptability (6): Opportunities to try new tools exist, but due to time pressure and competition feel most familiar ones are preferred \cite{10.1145/3472688.3472689}.
	
	\item Passion (6): Engagement is strong, though social issues often occur \cite{10.1145/3472688.3472693}.
	
	\item Versatility (7): Rapid prototyping encourages using diverse tools, dependent on personal initiative and team context \cite{10.1145/3472688.3472689}.
\end{itemize}

\subsubsection{Indie Development}
\begin{itemize}
	\item Skill Evaluation (10): Working solo or in very small teams pushes individuals to their limits; failure has immediate consequences, providing intense learning and skill validation. \cite{freeman2019}
	
	\item Credibility (8): Success can earn strong recognition, particularly within niche communities, though visibility is limited among the broader industry \cite{su2023}.
	
	\item Sustainability (3): High risk and volatility; income is unpredictable, and maintaining output requires mastery or exceptional luck. Some direct consumer rewards exist, but stability is rare \cite{wifitalents2025, freeman2023, su2023}.
	
	\item Adaptability (7): Indie developers frequently adjust to new tools and workflows, but it is more personal preference rather than formalized evaluation processes \cite{jgeekstudies2017}.
	
	\item Passion (9): High intrinsic motivation drives projects; autonomy allows personal creative freedom, though external support is limited \cite{freeman2019}.
	
	\item Versatility (8): Necessity to handle multiple roles encourages broad skill sets, often aligned with industry standards; however these skills may be incomplete or self-taught \cite{freeman2019}.
\end{itemize}

\subsubsection{Formal Education}
\begin{itemize}
	\item Skill Evaluation (7): Structured courses provide strong skill evaluation via assignments and exams \cite{connor2016}, but the assessment often reflects academic standards rather than real-world industry expectations \cite{mcgill2009}. Support from instructors reduces the need to struggle independently, which limits exposure to practical problem-solving \cite{educsci14111181}.
	
	\item Credibility (6): Degrees and school reputation can facilitate industry entry and networking, but they do not guarantee recognition or employment; personal initiative and portfolio quality remain crucial \cite{sign2023}. 
	
	\item Sustainability (7): Formal education, when appropriately focused, can enhance job prospects and earning potential \cite{czauderna2018}; however, high costs and delayed entry into the workforce can create financial precarity \cite{pixlvisn2025}.
	
	\item Adaptability (2): Curricula are still rigid and outdated; learners must follow predefined tools and methods, limiting the development of flexibility and responsiveness to industry changes \cite{milara2024}.
	
	\item Passion (4): Some motivation comes from collaborative learning and structured guidance \cite{10.1145/3725859}, but individual creative choice is constrained by curriculum requirements \cite{milara2024}.
	
	\item Versatility (6): Education provides a solid foundation and familiarity with industry-standard tools, but exposure to multiple engines or experimental workflows is limited \cite{sign2023}.
\end{itemize}

\subsubsection{Employment}
\begin{itemize}
	\item Skill Evaluation (7): Daily tasks in a studio require consistent skill application, though support from teammates reduces risk of complete failure. Exposure to complex challenges exists, but backup mechanisms moderate the pressure \cite{saud2024}.
	
	\item Credibility (9): In studios, professional recognition develops quickly, and this immediate visibility facilitates networking, making future opportunities more accessible \cite{espersson2024}. However, outside studio credibility can be dependent on organizational performance and hierarchy.
	
	\item Sustainability (9): Employment provides steady income and benefits. High performers may earn beyond standard rates, but less compared to other industries \cite{onet2025}.
	
	\item Adaptability (3): Tool and workflow choices are largely dictated by organizational constraints and licensing, limiting individual flexibility. Some adaptability exists, but it is mostly top-down \cite{cgcookies2023}.
	
	\item Passion (7): Collaborative, skilled teams and tangible end products enhance intrinsic motivation. Personal fulfillment may be limited by team composition, its dynamics or misaligned tasks \cite{alketbi2023, library994754}.
	
	\item Versatility (6): Creative workers can contribute to non-creative industries \cite{hearn2014}; however, efforts in specialization are often undervalued.
\end{itemize}

\subsubsection{Portfolio Projects}
\begin{itemize}
	\item Skill Evaluation (8): Portfolios push practitioners to demonstrate their best work and when aligned with meaningful goals, they can effectively test technical and creative skills but self-directed choices may lead to uneven challenge levels \cite{4417802}. 
	
	\item Credibility (7): Portfolios provide visible evidence of ability and can include studio or collaborative work \cite{watanabe2025}; however, individual contributions may be unclear, and compatibility with team-based work remains uncertain.
	
	\item Sustainability (3): Fully independent portfolio projects carry the risk of commercial non-viability. Even a portfolio career demands multiple income sources to become sustainable \cite{munnelly2022}.
	
	\item Adaptability (3): A portfolio is a curated showcase, projects are either self-directed or the evidence is selectively chosen, the results only prove success under controlled conditions, showing only preparations of highlights, not adaptions to the unexpected constraints.
	
	\item Passion (8): Practitioners generally pursue projects aligned with personal interest, resulting in motivated and creative output \cite{keune2017my, munnelly2022}; however the excessive external evaluation can undermine the motivation \cite{amabile1979}.  
	
	\item Versatility (9): Although a portfolio can effectively showcase command of distinct techniques and styles, the individual must choose to display that range \cite{munnelly2022}.
\end{itemize}

\subsubsection{Tutorials}
\begin{itemize}
	\item Skill Evaluation (1): Tutorials primarily instruct on how to solve a problem, rather than evaluate the learner’s skills.
	
	\item Credibility (3): Tutorial work is easily replicated and therefore lacks the uniqueness required to establish strong professional credibility.
	
	\item Sustainability (3): Low barrier to entry and lack of uniqueness limit market value.
	
	\item Adaptability (10): Tutorials offer on-demand, modular learning perfect for rapidly changing tech fields.
	
	\item Passion (8): High intrinsic motivation but constrained by existing content limits. Learners pursue content aligned with personal interests, but are also limited by what is available.
	
	\item Versatility (6): While tutorials are often task and tool-specific, a learner can achieve versatility by synthesizing and transferring methods; this process requires extra effort.
\end{itemize}

\subsection{Quantitative Analysis of existing Methods}

Using the qualitative scoring from the previous section, each method was evaluated across six requirements: Skill Testing, Credibility, Sustainability, Adaptability, Passion, and Versatility in the following Table: 

\begin{table}[h!]
	\centering
	\scriptsize
	\caption{Qualitative scoring of established creative technology methods across key requirements (0 = not promoting, 10 = strongly promoting).}
	\begin{tabular}{lcccccc}
		\toprule
		\textbf{Method} & \textbf{Skill Evaluation} & \textbf{Credibility} & \textbf{Sustainability} & \textbf{Adaptability} & \textbf{Passion} & \textbf{Versatility} \\
		\midrule
		Game Jams           & 8  & 7  & 3  & 6  & 6  & 7  \\
		Indie Development   & 10 & 8  & 3  & 7  & 9  & 8  \\
		Formal Education    & 7  & 6  & 7  & 2  & 4  & 6  \\
		Employment          & 7  & 9  & 9  & 3  & 7  & 6  \\
		Portfolio Projects  & 8  & 7  & 3  & 3  & 8  & 9  \\
		Tutorials           & 1  & 3  & 3  & 10 & 8  & 6  \\
		\midrule
		\textbf{Average}    & 6.83 & 6.67 & 4.67 & 5.17 & 7.0 & 7.0 \\
		\bottomrule
	\end{tabular}
	\label{tab:methods_scores}
\end{table}

\begin{figure}[h]
  \centering
  \includegraphics[width=\linewidth]{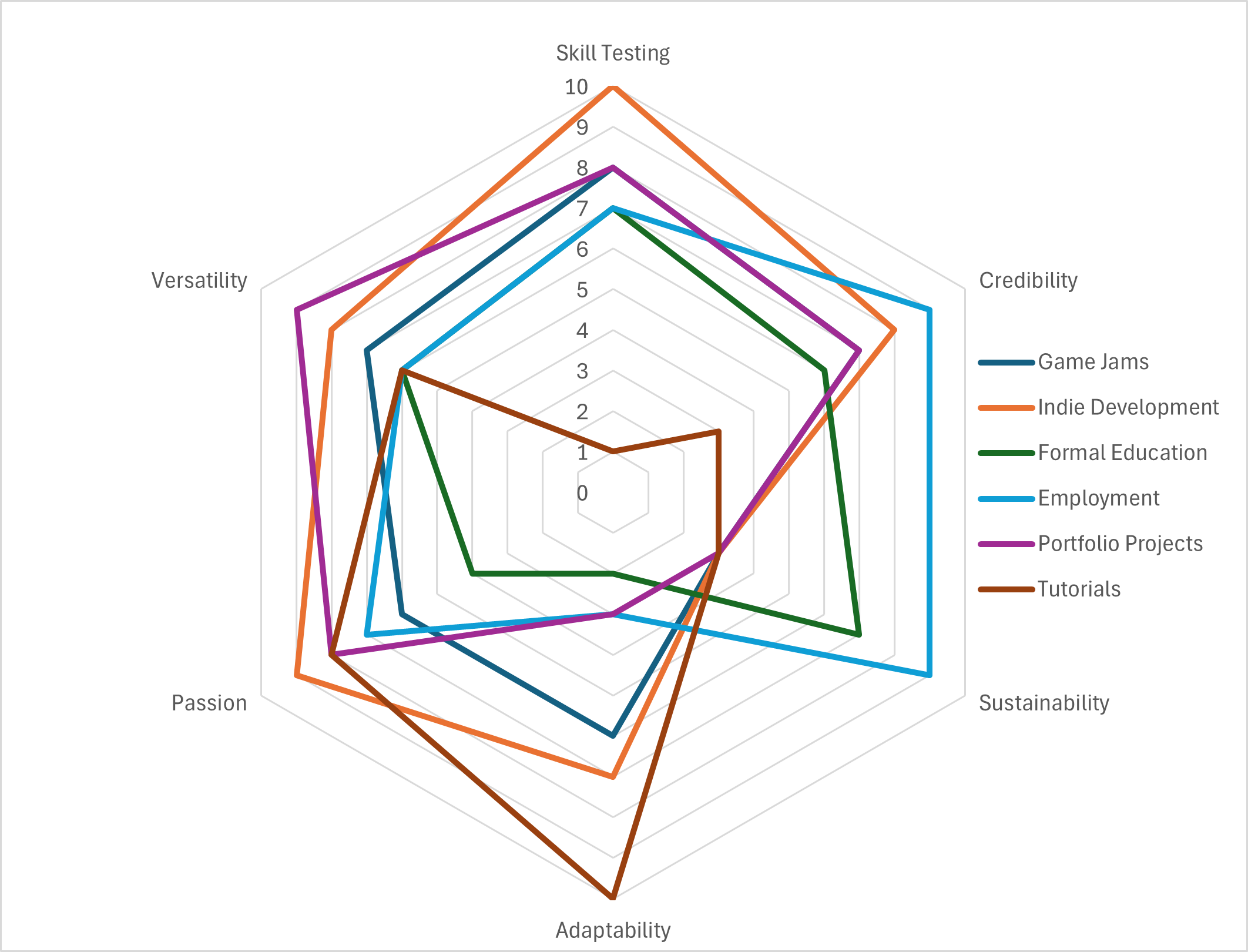}
  \caption{Plot of Qualitative Scoring of established methods across key requirements (0 = not promoting, 10 = strongly promoting).}
  \Description{Scores range from 0 (not promoting) to 10 (strongly promoting).}
  \label{fig:radarplot_scores}
\end{figure}

The qualitative scoring results are visualized in the radar plot ~\ref{fig:radarplot_scores}, revealing a noticeable gap between adaptability and sustainability across established methods.

\subsection{Observations}
Based on the quantitative summary, Indie Development achieves the highest total score (45), driven primarily by its maximum Skill Evaluation (10) and Passion (9). Employment follows closely (41), standing out for credibility (9) and sustainability (9) but by poor adaptability (3). Across all methods, Sustainability (mean = 4.67) remains the weakest dimension, while Passion and Versatility (both 7.0) emerge as the strongest. The most balanced methods—those without extreme highs or lows—are Game Jams and Portfolio Projects, though their sustainability scores remain critically low (3), indicating a structural issue in long-term stability.

A clear imbalance appears between Adaptability and Sustainability: tutorials achieve extremely high adaptability (10) but remain unsustainable (3), while employment reverses this pattern with high sustainability but poor adaptability (3). This contrast highlights that current creative technology pathways tend to specialize narrowly—either allowing fast learning and freedom or offering financial stability—but rarely both.

In practice, individuals seldom engage with all six methods simultaneously. Each path demands substantial time, focus, and effort, and combining many produces diminishing returns. Humans exhibit limited capacity for effective multitasking \cite{ophir2009}, which increases the risk of overload when managing multiple creative venues. Therefore, practitioners typically pursue one or two paths—such as employment plus a side project—rather than juggling all methods. This limitation amplifies existing weaknesses: for example, relying primarily on game jams or tutorials yields high adaptability but exposes one to financial and sustainability risks, while focusing solely on employment secures stability but limits creative growth and responsiveness to new technologies.

Even if, in theory, the entire ecosystem of methods would cover all requirements, none can leverage them all effectively. The result is an uneven distribution of strengths across the industry, reflected in the broader perception of game development as volatile, underpaid, and unstable.

To address this, the proposed framework unites learning and professional practice within a single, continuous model. The learning phase functions as adaptable while the professional phase forms the sustainable pillar.

The dynamic interface between skill acquisition and sustainable professional engagement must remain resilient under conditions of uncertainty. In the following proposed model, completion is not regarded as a terminal point but as a foundation for continuous iteration and growth, allowing both individuals and the broader industry to evolve in a healthy manner.

\section{The Framework}
The analysis of existing methods showed that, individually or combined, they generally support Skill Testing, Credibility, Passion, and Versatility. However, the adaptability–sustainability corner is underrepresented, creating risk for both learners and professionals.
This shortcoming arises from the gap between education and profession: Adaptability is cultivated through education, and sustainability through profession, yet because these systems evolved apart, their interdependence remains underdeveloped. Education and profession must operate in tandem so that adaptability and sustainability can connect. 

\begin{figure}[h]
	\centering
	\includegraphics[width=\linewidth]{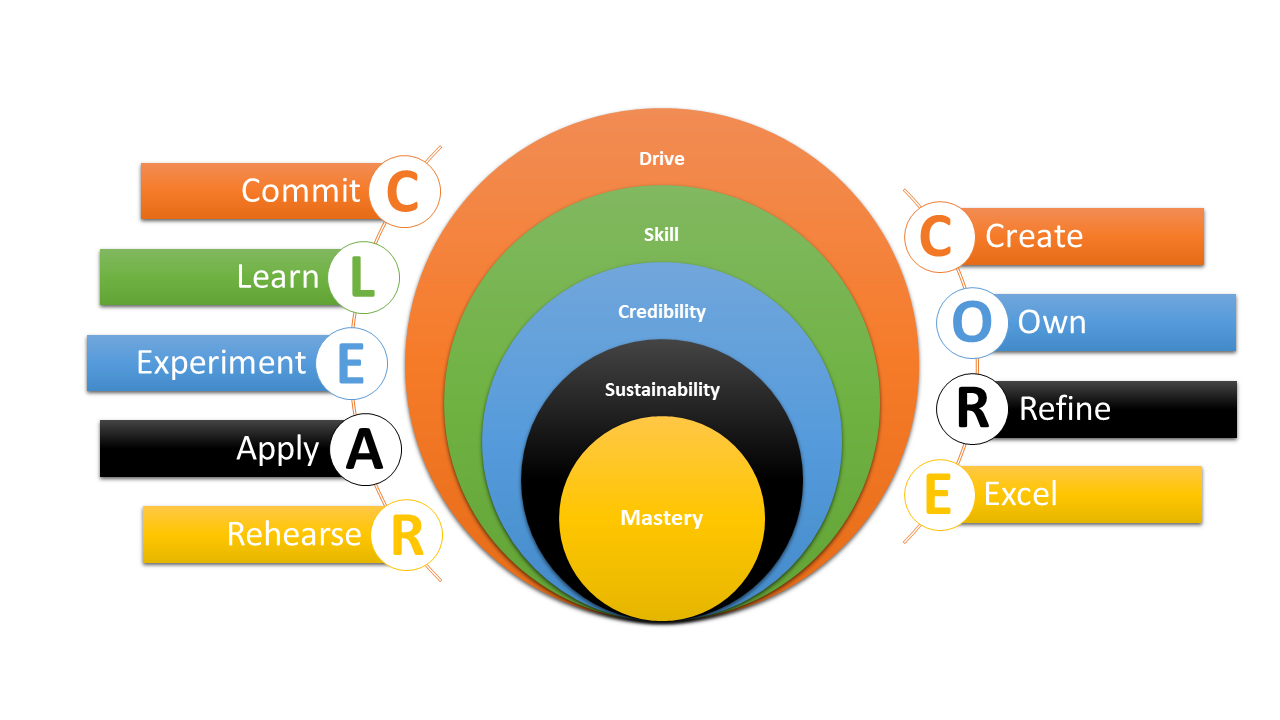}
	\caption{The Creative Technology Framework CLEAR CORE}
	\Description{left to right: CLEAR (Education), The Outcome, CORE (Profession)}
	\label{fig:framework}
\end{figure}

To make this connection, the proposed CLEAR CORE framework, visualized in figure ~\ref{fig:framework}, establishes an integrated model that unifies education and profession—that is, learning and working—into a continuous developmental loop. It does so through two interdependent columns: CLEAR (education) and CORE (profession). 

\begin{itemize}
	\item In CLEAR (education), adaptability is given through iterative experimentation, reflective learning, and engagement with evolving tools or approaches.
	\item In CORE (profession), sustainability is given by performing repeatable, professional workflows that produce reliable, meaningful outcomes over time.
\end{itemize}

The middle column of the framework—\textit{the Outcome}—represents a synthesis of CLEAR and CORE. It is the point where education and profession converge. 
\textit{The Outcome} uses integrated labels that reflect the framework's holistic nature: \textit{Drive} (from Passion), \textit{Skill} (from Skill Evaluation), \textit{Credibility}, \textit{Sustainability}, and \textit{Mastery} (covering both Adaptability and Versatility).
Although Sustainability and Adaptability were the recognized gaps in existing methods, they are separated into the top two layers because Sustainability serves as the foundation for Mastery, which then encompasses both the ability to adapt (Adaptability) and to apply skills broadly (Versatility). All five layers collectively address the previously identified six requirements: \textit{Passion}, \textit{Skill Evaluation}, \textit{Credibility}, \textit{Sustainability}, \textit{Adaptability}, \textit{Versatility}.

The Creative Technology Framework ensures that whatever method or combination of methods an individual engages with, adaptability and sustainability are inherently accounted for. All other requirements remain fulfilled, as they emerge organically from the granular application and combination of existing methods.

CLEAR, CORE, and Outcome are described in detail below.

\subsection{CLEAR}
The CLEAR column defines the \textit{educational front} of the framework—how creative technologists learn, reflect, and internalize skills through structured iteration. It provides a rhythm for growth that moves beyond the accumulation of knowledge toward transformation through practice. CLEAR stands for the following stages and establishes a self-reinforcing cycle of adaptive learning.

\begin{itemize}
	\item \textbf{Commit} represents intentional engagement: setting a clear purpose, defining what and why to learn.
	\item \textbf{Learn} encompasses acquisition of knowledge and techniques through study, observation and/or mentorship.
	\item \textbf{Experiment} invites exploration through trials—testing limits and failing safely to learn effectively.
	\item \textbf{Apply} converts experimentation into practice, embedding insights into meaningful projects or workflows.
	\item \textbf{Rehearse} ensures retention and refinement through repetition and reflection.
\end{itemize}

CLEAR is designed to balance curiosity with discipline. It validates growth through evidence of progress, not perfection. By encouraging deliberate cycles of exploration and application, it nurtures learners who are not only skilled but adaptable—capable of navigating evolving technologies and shifting creative landscapes.

\subsection{CORE}
The \textbf{CORE} column defines the \textit{professional front}—how creative technologists work, deliver, and sustain excellence in real-world environments. While CLEAR focuses on learning, CORE translates those learnings into professional capability. It represents the continuous refinement of creative and technical performance through accountability, collaboration, and impact. CORE stands for the following, forming an iterative cycle of professional mastery.

\begin{itemize}
	\item \textbf{Create} focuses on producing meaningful, high-quality outputs that serve a purpose or audience.
	\item \textbf{Own} emphasizes originality and personal value of the work.
	\item \textbf{Refine} represents the continuous pursuit of improvement, addressing feedback and optimizing workflows.
	\item \textbf{Excel} embodies professional resilience and distinction—its achievement is a consistent standard of practice.
\end{itemize}

CORE challenges expertise through consistent delivery and integrity under real constraints. Together, these actions cultivate a professional rhythm—an equilibrium between creative autonomy and industry expectations, ensuring practitioners remain both independent and compatible with collaborative production environments.

\subsection{Outcome}
The Outcome column is the structured progression through which education (CLEAR) and profession (CORE) converge into enduring capability. It consists of five hierarchical layers: Drive, Skill, Credibility, Sustainability, and Mastery. Each layer is both a foundation and a support for the one above it, meaning the size and strength of the lower layers must exceed that of the upper layers to maintain structural integrity.

\begin{itemize}
	\item \textbf{Drive} forms the foundation and is the largest layer. It embodies intrinsic motivation, curiosity, and perseverance. Without sufficient Drive, none of the upper layers can sustain themselves. Drive fuels any involved action.
	
	\item \textbf{Skill} rests on Drive. It represents the tangible competencies, the practitioner develops, recognizing that significant Drive is required for any skill acquisition.
	
	\item \textbf{Credibility} builds atop Skill. It represents external recognition, trust, and professional reliability, which must be earned by cultivating a depth of Skill beyond immediate credit. This foundation of Skill and Drive allows practitioners to demonstrate competence in real-world contexts.
	
	\item \textbf{Sustainability} emerges as Credibility solidifies, representing the ability to maintain high-quality work over time. It balances ambition, health, and adaptability, ensuring continued relevance and productivity.
	
	\item \textbf{Mastery} crowns the ladder as the smallest, most refined layer. It represents the synthesis of Drive, Skill, Credibility, and Sustainability into almost effortless, recognized fluency. Even if Mastery diminishes, the underlying layers—particularly Drive—allow the practitioner to rebuild and adapt without collapsing the entire framework.
	This layer also encompasses Adaptability and Versatility. It depends on Sustainability because without sustained engagement in practice, flexibility and adaptability degrade over time.
\end{itemize}

This layered structure emphasizes that Drive must exceed Skill, Skill must exceed Credibility, and so on, because each layer not only supports the next but also defines a smaller capacity of the layers above. The Outcome column thus translates the dual efforts of CLEAR and CORE into a resilient, self-supporting system: robust at its base, capable of growth, and most fulfilling at its apex.

\subsection{Interconnections}

The strength of the proposed framework lies not only in the distinct roles of CLEAR and CORE, but also in the dynamic interconnections that link educational processes to professional practice. These connections ensure that learning and work are mutually reinforcing, enabling sustained growth and resilience.
Following shows most significant interconnections:
\subsubsection{Drive $\rightarrow$ Commit $\leftrightarrow$ Create}
	
	In CLEAR, practitioners \textit{commit} to a learning path and dedicate themselves to growth.  
	In CORE, they \textit{create} by transforming that learning into tangible outcomes.  
	The link between the two is bidirectional: creation fuels further commitment, and commitment enables meaningful creation.  
	\textit{Drive} thus becomes the shared energy that unites both columns in a continuous cycle of learning and doing.
	
\subsubsection{Skill $\rightarrow$ Learn}
	
	The stage \textit{Learn} in CLEAR, involving structured study, reflection, and comprehension, maps directly to the \textit{Skill} in the outcome. The knowledge and abilities acquired through learning become the foundation of technical and creative proficiency. Without effective learning, skill growth is limited; conversely, active skill application reinforces and deepens the learning process.

\subsubsection{Credibility $\rightarrow$ Experiment $\leftrightarrow$ Own}

To \textit{experiment} in CLEAR fosters creative exploration and iterative testing. When paired with the CORE practice of taking \textit{ownership} over work, experimentation translates into tangible professional output that earns industry recognition. \textit{Credibility} emerges as learners validate their experiments in real-world workflows, demonstrating both technical competence and professional reliability.

\subsubsection{Sustainability $\rightarrow$ Apply $\leftrightarrow$ Refine}

To \textit{apply} in CLEAR, where learned skills are put into practice, connects to \textit{refine} in CORE, where those skills are improved through professional execution and feedback. This iterative loop enables \textit{sustainability}, as practitioners must continuously \textit{apply} and \textit{refine} their skills to maintain competence and relevance within evolving professional contexts.

\subsubsection{Mastery $\rightarrow$ Rehearse $\leftrightarrow$ Excel}
To \textit{rehearse} in CLEAR, where techniques are refined through repetition, connects to \textit{excel} in CORE, where \textit{mastery} is demonstrated through sustained excellence. To \textit{excel} and \textit{rehearse} are the acts that keeps \textit{mastery} alive.

These interconnections are not rigid mechanisms but conceptual guides, illustrating how elements influence and reinforce one another. They do not operate like screws and gears that engage in a strict sequence. Instead, the model functions like an orchestrated lifestyle: there is no single direction or linear path, and no fallback point. Each layer and element continuously affects the others, creating a holistic, living system where learning, professional application, and growth feed into one another.

\section{Framework Effectiveness: Scenario-Based Validation}

The CLEAR CORE framework addresses the professional challenges of creative technologists by providing dual fronts of support—educational and professional—and producing a layered outcome of Drive, Skill, Credibility, Sustainability, and Mastery. The following scenarios illustrate how the framework would function in concrete situations:

\subsection{Scenario 1: Layoff at a Mid-Size Game Studio}

\paragraph{\textbf{Case Description:}}
A character artist has built an extensive portfolio over several years, but nearly all work originates from team projects within the studio’s proprietary pipeline. When the studio undergoes restructuring and the artist is laid off, the portfolio demonstrates technical skill but no workflow independence or adaptability. Most prior experience is tightly coupled to the studio’s internal tools, asset rights, and stylistic conventions, limiting transferability.

\paragraph{\textbf{Typical Outcome:}}
Because the portfolio is so tightly coupled to the former studio’s pipeline, the artist struggles to find new opportunities. Other studios cannot readily use the showcased work (low adaptability), and the prior experience does not demonstrate sustainability. Professional continuity is compromised, and the artist faces an uncertain career path.

\paragraph{\textbf{CLEAR CORE Outcome:}}
Using the framework, the artist maintains a portfolio of personal, reusable assets and experiments with multiple methods and workflows (CLEAR). In professional practice (CORE), the artist refines these methods and demonstrates their applicability across contexts. When laid off, the artist can sell assets, take commissions, or pursue new studio roles with confidence. Recruiters see both adaptability and industry-relevant skills, and the practitioner maintains professional continuity and career resilience.

\subsection{Scenario 2: Technology Shift}

\paragraph{\textbf{Case Description:}}
Developer A has specialized for years in the studio’s current game engine. The studio considers switching to a different one, and prioritizes developers already familiar it.

\paragraph{\textbf{Typical Outcome:}}
Because A’s expertise is tied to the older engine, the studio assigns A to minor projects and provides no formal training. A misses opportunities to learn the new engine, cannot contribute to high-priority projects, and risks losing professional relevance.

\paragraph{\textbf{CLEAR CORE Outcome:}}
Developer A \textit{experiments} with the new engine version immediately upon release (CLEAR) and \textit{creates} a few unique demos (CORE). When the studio transitions, A is prepared to lead the migration of a new production pipeline, maintaining professional standing, adaptability, and long-term career sustainability.

\subsection{Scenario 3: Client-Driven vs. Passion Projects}

\paragraph{\textbf{Case Description:}}
A game designer works primarily on publisher-driven projects that demand strict adherence to external creative directions. Over time, these client requests begin to conflict with the designer’s personal creative goals, creating tension between professional obligations and intrinsic motivation.

\paragraph{\textbf{Typical Outcome:}}
The designer continues to meet client demands while postponing personal projects. After several years of producing similar content under varying publisher expectations, creative motivation declines. The designer’s portfolio reflects technical skill but lacks personal identity or originality. Eventually, they lose interest in the profession and struggle to re-engage creatively.

\paragraph{\textbf{CLEAR CORE Outcome:}}
The designer uses client projects as structured opportunities to \textit{learn} new tools, genres, and design systems (CLEAR) while dedicating time to \textit{create} and refine personal prototypes or experimental game ideas (CORE). Client work provides technical and financial stability, while personal projects sustain passion and creative development.  
Over time, the designer’s portfolio evolves to demonstrate both commercial reliability and a distinct creative voice. Clients begin to trust the designer’s direction, allowing more creative freedom in commissioned projects.

\subsection{Scenario 4: Entry-Level Graduate vs. Industry Expectations}

\paragraph{\textbf{Case Description:}}
A recent game development graduate from a four-year program applies for multiple positions and internships across the gaming industry, hoping to begin professional work immediately after graduation. The graduate has student loan debt exceeding \$25,000 and feels strong pressure to secure employment quickly. During the program, a portfolio of solo capstone projects was developed in Unreal Engine using university-provided templates.

\paragraph{\textbf{Typical Outcome:}}
The graduate’s projects demonstrate technical understanding, but recruiters describe the portfolio as “too academic.” The showcased work lacks evidence of collaboration, code review, or production-scale iteration. Version control history shows no team contributions, and gameplay systems are hard-coded without modular design. After multiple rejections, the graduate struggles financially and considers leaving the game industry for a more stable software role to repay debt.

\paragraph{\textbf{CLEAR CORE Outcome:}}
During the program, the graduate joins public GitHub projects with daily collaboration and structured issue tracking to gain hands-on experience with professional workflows. They also participate in several Unreal Engine game jams, focusing on creating small but releasable prototypes.
After graduation, one of these game jam projects gains traction online, attracting the attention of a small startup studio. Impressed by the graduate’s demonstrated teamwork and technical skill, the studio offers an internship that develops into a full-time role within a year. Through this position, the graduate quickly repays most of their student debt and establishes a reputation as a dependable gameplay programmer.

\subsection{Scenario 5: Mid-Career Plateau / Creative Fatigue}

\paragraph{\textbf{Case Description:}}
A senior technical artist has been working at the same studio for over a decade. Their role is stable but repetitive, involving incremental improvements rather than innovation. Creative fatigue sets in, and professional growth slows.

\paragraph{\textbf{Typical Outcome:}}
Without deliberate learning or creative renewal, the artist’s motivation declines. They deliver consistent but uninspired work, become resistant to change, and risk obsolescence as new technologies and techniques emerge.

\paragraph{\textbf{CLEAR CORE Outcome:}}
Through CLEAR, the artist deliberately schedules learning cycles to explore emerging technologies, such as procedural workflows or AI-assisted tools. In CORE, these experiments are applied to internal production optimizations or side prototypes that increase pipeline efficiency. The renewed engagement rekindles passion and strengthens long-term sustainability.

\subsection{Rationale}
The CLEAR CORE framework is a proactive professional mindset, not a reactive remedy. Applicable at any career stage, its true strength lies in anticipation: it cultivates the necessary elements proactively to ensure long-term growth.

The scenarios presented illustrate how professionals who embody the framework maintain relevance and continuity even through industry shifts, layoffs, or early-career challenges. Though only a selection is shown here, each is grounded in real industry experiences and reflects common situations within creative technology disciplines. The underlying principle remains constant: with CLEAR CORE, adaptability and sustainability are continuously reinforced—preventing problems before they arise.

\section{Conclusion}

The CLEAR CORE framework successfully bridges the gap between learning and professional practice for fields depending on computing education. It utilizes dual, interconnected columns for education (CLEAR) and professional application (CORE), leading to a synthesized Outcome layer encompassing Drive, Skill, Credibility, Sustainability, and Mastery that ensures continuous growth and resilience. Initial examples validated the model's efficacy, demonstrating the closure of the critical gaps in sustainability and adaptability while satisfying all other requirements. The framework’s core achievement is that it ensures all requirements are consistently met—regardless of the practitioner’s chosen method—provided the guiding elements of CLEAR CORE are followed. This ensures that individuals with a computing education background—particularly creative technologists who encounter the greatest challenges—can pursue their passion, achieve professional excellence, and sustain long-term career viability and satisfaction. While this study provides foundational validation, future research will focus on additional rigorous, quantitative, and comparative longitudinal studies across diverse methodologies to provide deeper, empirical evidence of the framework's impact in education and professional settings.

\bibliographystyle{ACM-Reference-Format}
\bibliography{references}

\appendix
%
%
%
%
%
%
%
%

\end{document}